\documentclass[prl,groupedaddress,superscriptaddress,twocolumn,showpacs,amsmath,amssymb,a4paper]{revtex4}
\usepackage{geometry}
\usepackage{graphicx}
\usepackage{verbatim}
\usepackage{float}
\usepackage{graphics}
\usepackage{mathrsfs}
\usepackage{amssymb}
\usepackage{amsmath}
\usepackage{amsthm}
\usepackage{bm}
\usepackage{dsfont}
\usepackage{hyperref}
\usepackage{braket}

\usepackage{amsbsy}

\theoremstyle{definition}

\newcommand{\ip}[2]{{\langle #1|}{ #2 \rangle }}

\newcommand{\be}{\begin{eqnarray}}
\newcommand{\ee}{\end{eqnarray}}

\begin{document}

\title{Simulation complexity of open quantum dynamics: Connection with tensor networks}

\author{I. A. Luchnikov}
\affiliation{Center for Energy Science and Technology, Skolkovo Institute of Science and Technology, 3 Nobel Street, Skolkovo, Moscow Region 121205, Russia}
\affiliation{Moscow Institute of Physics and Technology, Institutskii Per. 9, Dolgoprudny, Moscow Region 141700, Russia}

\author{S. V. Vintskevich}
\affiliation{Moscow Institute of Physics and Technology, Institutskii Per. 9, Dolgoprudny, Moscow Region 141700, Russia}
\affiliation{A.M. Prokhorov General Physics Institute, Russian Academy of Sciences, Moscow, Russia}

\author{H. Ouerdane}
\affiliation{Center for Energy Science and Technology, Skolkovo Institute of Science and Technology, 3 Nobel Street, Skolkovo, Moscow Region 121205, Russia}

\author{S. N. Filippov}
\affiliation{Moscow Institute of Physics and Technology,
Institutskii Per. 9, Dolgoprudny, Moscow Region 141700, Russia}
\affiliation{Valiev Institute of Physics and Technology of Russian
Academy of Sciences, Nakhimovskii Pr. 34, Moscow 117218, Russia}
\affiliation{Steklov Mathematical Institute of Russian Academy of
Sciences, Gubkina St. 8, Moscow 119991, Russia}

\begin{abstract}
The difficulty to simulate the dynamics of open quantum systems resides in their coupling to many-body reservoirs with exponentially large Hilbert space. Applying a tensor network approach in the time domain, we demonstrate that effective small reservoirs can be defined and used for modeling open quantum dynamics. The key element of our technique is the timeline reservoir network (TRN), which contains all the information on the reservoir's characteristics, in particular, the memory effects timescale. The TRN has a one-dimensional tensor network structure, which can be effectively approximated in full analogy with the matrix product approximation of spin-chain states. We derive the sufficient bond dimension in the approximated TRN with a reduced set of physical parameters: coupling strength, reservoir correlation time, minimal timescale, and the system's number of degrees of freedom interacting with the environment. The bond dimension can be viewed as a measure of the open dynamics complexity. Simulation is based on the semigroup dynamics of the system and effective reservoir of finite dimension. We provide an illustrative example showing scope for new numerical and machine learning-based methods for open quantum systems.
\end{abstract}

\pacs{02.10.Xm,02.10.Yn,03.65.Yz,03.65.Aa,03.65.Ca}

\maketitle

\paragraph*{Introduction.}
One of the most challenging and important problems of modern
theoretical physics is the accurate simulation of an interacting
many-body system. As the dimension of its Hilbert space grows
exponentially with the system size, direct simulations become
impossible. Exactly solvable models exist nonetheless
\cite{Sutherland2004}; they provide some insights into the
properties of actual physical systems. Perturbation theory can be
used only for problems that can be split into an exactly solvable
part and a perturbative one provided that a relevant small
parameter (e.g., weak interaction strength with respect to other
energy scales) can be defined. For strongly interacting many-body
systems, a range of techniques including, e.g., the Bethe ansatz
\cite{Bethe1931}, the dynamical mean field theory
\cite{Georges1996,Kotliar2006}, or the slave boson techniques
\cite{Barnes1976,Kotliar1986} have been developed and applied to
problems like the diagonalization of the Kondo Hamiltonian or the
Anderson impurity model
\cite{Andrei1980,Wiegmann1980,Georges1992,Ouerdane2007,Ouerdane2008}.
Numerical approaches, which may significantly go beyond the range
of applicability of analytical methods, have been also developed
and proved quite successful, though they suffer from limits. For
example, methods based on tensor networks
\cite{Orus2014,Orus2014b,Verstraete2008} and the density matrix
renormalization group
\cite{White1992,Schollwock2011,Schollwock2011b} work well mainly
for one-dimensional models. Quantum Monte Carlo (QMC) methods
provide reliable ways to study the many-body problem
\cite{Landau2009}, but for interacting fermion systems these
approaches are plagued by the sign problem \cite{Loh1990}.

\begin{figure}
\includegraphics[width=8cm]{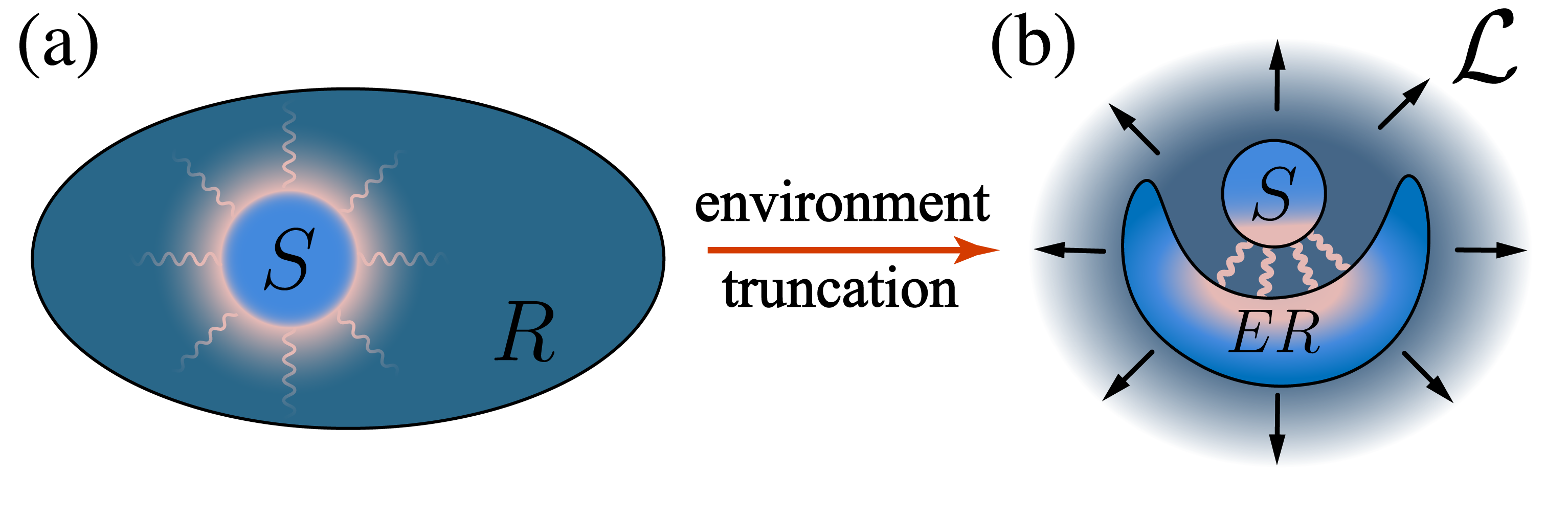}
\caption{Schematic of reservoir truncation.} \label{fig1}
\end{figure}

Unitary evolution of a many-body system is completely out of
reach: dynamical versions of QMC calculations or efficient methods like 
the time-evolving block decimation algorithm cannot predict long-term
time dynamics because of the Lieb-Robinson bound
\cite{Lieb1972,prosen-2007,schollwock-2013,InchWorm}. Having
experimental access only to a part of a many-body system, one in
fact deals with an {\it open} quantum dynamics of the subsystem
($S$), whereas the rest of the particles (modes) play the role of
environment also referred to as reservoir ($R$), see
Fig.~\ref{fig1}(a). The subsystem is described by a density
operator $\rho_S(t) = {\rm tr}_{R}[U(t) \rho(0) U^{\dag}(t)]$, the
evolution of which is still challenging to determine though the
subsystem is relatively small compared to the
environment~\cite{Davies1976,Breuer2002}: the partial trace, ${\rm
tr}_R$, disregards the environment degrees of freedom but
$\rho(0)$ is the initial state of the \emph{whole} many-body
system and its evolution operator is $U(t) = e^{-itH}$. There
exist particular exactly solvable models of open quantum
dynamics~\cite{rau-1963,palma-1996,fpmz-2017}; however, without
the Markov approximation the problem of open
dynamics is typically impossible to solve directly because of the
exponentially large dimension of the reservoir's Hilbert
space~\cite{Breuer2002,Alicki2012,Holevo2012}. Examples of complex
open dynamics in structured reservoirs, where it is necessary to
go beyond the Markov approximation, are presented
in~\cite{piilo-2011,cirac-2011,ma-2012,hoope-2012,yang-2013,hughes-2015,eisert-2015,cirac-2017,wittemer-2018,wang-2018,peng-2018,haase-2018}.
Therefore, new, numerically tractable approaches permitting
significant progress in the field of open quantum dynamics
simulation are highly desirable, and their development
constitutes a timely challenge, especially in the study
of quantum control and dynamical decoupling~\cite{wang-2018,peng-2018,haase-2018},
and quantum dynamics induced by many-body reservoirs~\cite{pineda-2011,viyuela-2012,vasseur-2015,marzolino-2017}.

In this work, we show that the actual infinite environment can be
replaced by a \emph{finite}-dimensional effective reservoir ($ER$)
in such a way that the aggregate ``$S+ER$'' experiences semigroup
dynamics, see Fig.~\ref{fig1}(b). This approach resembles the idea
of Markovian embedding of non-Markovian
dynamics~\cite{budini-2013,xue-2015,xue-2017,tamascelly-2018} and
the pseudomode
method~\cite{imamoglu-1994,garraway-1997,mazzola-2009}. Our main
result is the estimation of the minimal (sufficient)
dimension $d_{ER}$ of the effective reservoir expressed through a
reduced set of parameters. Knowledge of $d_{ER}$ enables one to
efficiently simulate the complex dynamics of a subsystem of
dimension $d_S$ via $\rho_S(t) = {\rm tr}_{ER}[ e^{{\cal L}t}
\rho_{S+ER}(0) ]$, where the Gorini-Kossakowski-Sudarshan-Lindblad
(GKSL) generator ${\cal L}$~\cite{gks-1976,lindblad-1976} is easy
to parameterize in this case: ${\cal L}$ acts on $d_S d_{ER}
\times d_S d_{ER}$ matrices and can be numerically found via
machine learning techniques provided a sequence of measurements on
the subsystem is performed~\cite{luchnikov-ml-2019}. Another
machine learning algorithm~\cite{shrapnel-2018} estimates $d_{ER}$
within the training range $d_{ER} = 1,2,8,16$ based on
interventions in the open qubit evolution at 4 time moments.

\paragraph*{Open quantum systems properties.}
Let ${\cal H}_S$ and ${\cal H}_R$ be Hilbert spaces of the subsystem and the reservoir, respectively. Typically ${\rm dim}({\cal H}_S) \ll {\rm dim}({\cal H}_R)$ as ${\cal H}_S$ could be associated with, e.g., a qubit or other small system, and ${\cal H}_R$ with a many-body quantum environment of a huge dimension. The total Hilbert space is ${\cal H}={\cal H}_S\otimes{\cal H}_R$. As the environment is assumed to be in the thermodynamic limit, the dynamics of $\rho_S(t)$ is irreversible, i.e. the Poincar\'e recurrence time is infinite. When the subsystem and reservoir exchange energy, thermalization is expected on a long timescale~\cite{Kosloff2013}, though the dynamics can be strongly non-Markovian at finite times~\cite{de-vega-2017}.

The total Hamiltonian reads $H=H_0+H_{\rm int}$, where $H_0 =
H_S\otimes {\mathds 1}+{\mathds 1}\otimes H_R$ involves individual
Hamiltonians of the subsystem and the reservoir, $H_{\rm int}=\gamma
\sum_{i=1}^{n} A_i \otimes B_i$ is the interaction part with
characteristic interaction strength $\gamma$; $n$ is the effective
subsystem's number of degrees of freedom interacting with the
reservoir. Denote $B_i(t) = U^{\dag}(t) B_i U(t)$, then
$g_{ij}(t,t-s) = {\rm tr}[B_i^{\dag}(t) B_j(t-s) \rho(0)]-{\rm
tr}[B_i^{\dag}(t)\rho(0)]{\rm tr}[B_j(t-s) \rho(0)]$ is the
reservoir correlation function~\cite{arnoldus-1987}. Suppose
$g_{ij}(t,t-s)$ decays exponentially with the growth of $s$ over a
characteristic time $s_{ij}$ (see examples
in~\cite{weiss,eckel-2006}), then $T = \max_{ij} s_{ij}$ is the
reservoir correlation time. Suppose the Fourier transform of the
reservoir correlation function decays significantly at the
characteristic frequency $\Omega_{ij}$, then $\tau = (\max_{ij}
\Omega_{ij})^{-1}$ defines the minimal time scale in the dynamics.
In the case of bosonic bath, $\tau = \omega_c^{-1}$, where
$\omega_c$ is the cutoff frequency of the spectral
function~\cite{strathearn-2018}. Our approach to determine the
dimension $d_{ER}$ of truncated environment is based on tensor
network formalism in time domain, where the building blocks are
responsible for evolution during time $\tau$ and the ancillary
space is capable of transferring temporal correlations for the period
$T$. Therefore, $d_{ER}$ depends only on the following few
physical parameters: $\gamma$, $n$, $\tau$, and $T$.

\paragraph*{Tensor network representation of open quantum dynamics.}
For simplicity, we resort to the vector representation of a density operator:
\begin{equation}
\rho = \sum_{jk}\rho_{jk}\ket{j}\bra{k}\ \rightarrow \ \ket{\rho}=\sum_{jk} \rho_{jk} \ket{j} \otimes \ket{k},
\end{equation}

\noindent which implies $Q\rho P \rightarrow Q\otimes P^T\ket{\rho}$. The dynamics of the whole system reads
\begin{eqnarray}
\ket{\rho(t)} = \exp[-i t H] \otimes \exp[i t H^T]\ket{\rho (0)}.
\end{eqnarray}

\noindent The initial state $\rho(0)$ can be correlated in general, i.e. $\rho(0) = \sum_{l} \sigma_S^{(l)} \otimes \sigma_R^{(l)}$. The subsystem state ${\rho_S(t)} = {\rm tr}_R \rho(t)$ in terms of vectors reads $\ket{\rho_S(t)} = \bra{\psi_+} \ket{\rho(t)}$, where $\bra{\psi_+} = \sum_{j=1}^{d_{R}} \mathds{1}_{S}\otimes \bra{j}\otimes \mathds{1}_{S}\otimes \bra{j}$. For further convenience we introduce a new order of Hilbert spaces: $\cal{H_S}\otimes\cal{H_R}\otimes\cal{H_S}^\dagger\otimes\cal{H_R}^\dagger\rightarrow\cal{H_S}\otimes\cal{H_S}^\dagger\otimes\cal{H_R}\otimes\cal{H_R}^\dagger$.

The minimal timescale $\tau$ is a time step in discretized
evolution; $\tau$ can always be reduced in such a way that $\gamma
\tau \ll 1$, which permits application of the Trotter
decomposition~\cite{Suzuki1985}. Note that we do not restrict our
framework to a weak coupling between a subsystem and reservoir
($\gamma \ll \|H_0\|)$; we rather adjust the minimal timescale
$\tau$ in accordance with the interaction strength, which yields
\begin{equation}\label{rhoTrotter}
\ket{\rho(t)} = \underbrace{\Phi_0(\tau) \Phi_{\rm int} (\tau)
\cdots \Phi_0(\tau) \Phi_{\rm int}(\tau)}_{t/\tau~\text{times}}
\ket{\rho(0)} + O(\gamma\tau),
\end{equation}

\noindent where $\Phi_0(\tau)$ and $\Phi_{\rm int}(\tau)$ are responsible for the noninteractive and interactive evolutions of the subsystem and environment:
\begin{eqnarray}
\Phi_0(\tau) &=& \exp{[-i\tau H_{S}]} \otimes \exp{[i\tau H_{S}^T]} \nonumber\\
&& \otimes \exp{[-i\tau H_{R}]} \otimes \exp{[i\tau H_{R}^T]}, \\
\Phi_{\rm int}(\tau) &=& \sum_{i=0}^{2n} {\cal A}_i(\tau) \otimes
{\cal B}_i(\tau), \label{Phi-int} \\
{\cal A}_i(\tau) &=& \left\{ \begin{array}{ll}
  {\mathds 1}\otimes {\mathds 1} & \text{if~} i=0, \\
  \sqrt{\gamma\tau} \, A_i \otimes {\mathds 1} & \text{if~} 1 \leq i \leq n, \\
  \sqrt{\gamma\tau} \, {\mathds 1}\otimes A_{i-n}^T & \text{if~} i \geq n+1, \\
\end{array} \right. \label{1} \\
{\cal B}_i(\tau) &=& \left\{ \begin{array}{ll}
  {\mathds 1}\otimes {\mathds 1} & \text{if~} i=0, \\
  -i\sqrt{\gamma\tau} \, B_i \otimes {\mathds 1} & \text{if~} 1 \leq i \leq n, \\
  i\sqrt{\gamma\tau} \, {\mathds 1}\otimes B_{i-n}^T & \text{if~} i \geq n+1. \\
\end{array} \right. \label{1-1}
\end{eqnarray}

\begin{figure}
\includegraphics[width=8cm]{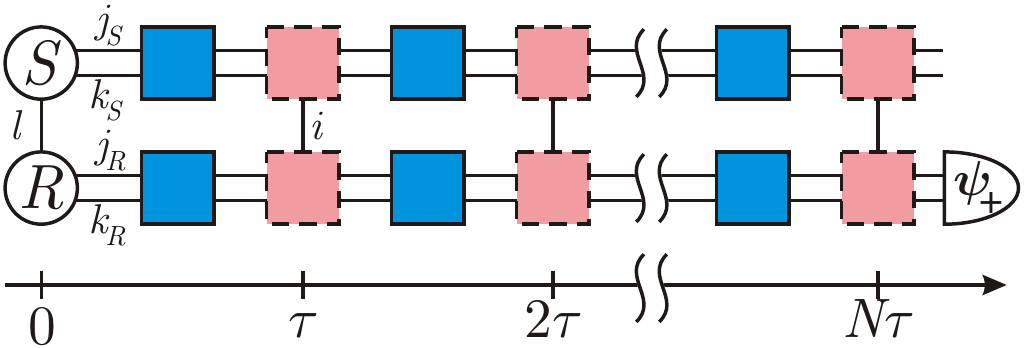}
\caption{Tensor network for open system dynamics.} \label{figure2}
\end{figure}

The tensor network representation to calculate $\ket{\rho_S(t)}$
is presented in Fig.~\ref{figure2}. It is a particular case of the
general quantum
circuits~\cite{chiribella-2008,chiribella-2009,hardy-2012,costa-2016,milz-2017}.
Each building block with $m$ arms corresponds to a tensor of rank
$m$. Connecting links denote contractions over the same indices.
The vector $\ket{\rho(t)}$ has two multi-indices $j=(j_S,j_R)$ and
$k=(k_S,k_R)$, so it is represented as a tensor of rank 4. The
upper (bottom) row corresponds to the degrees of freedom of
subsystem $S$ (reservoir $R$). The operator $\Phi_0(\tau)$ is
depicted by solid squares. The dashed squares with a link between
them denote the operator $\Phi_{\rm int}(\tau)$, with the link
being responsible for summation $\sum_{i=0}^{2n}$ in
formula~\eqref{Phi-int}. Concatenation with the building block
$\psi_+$ in right bottom of Fig.~\ref{figure2} corresponds to the
partial trace over $R$ and will be further denoted by connecting
link $\supset$.

\begin{figure}[b]
\includegraphics[width=8cm]{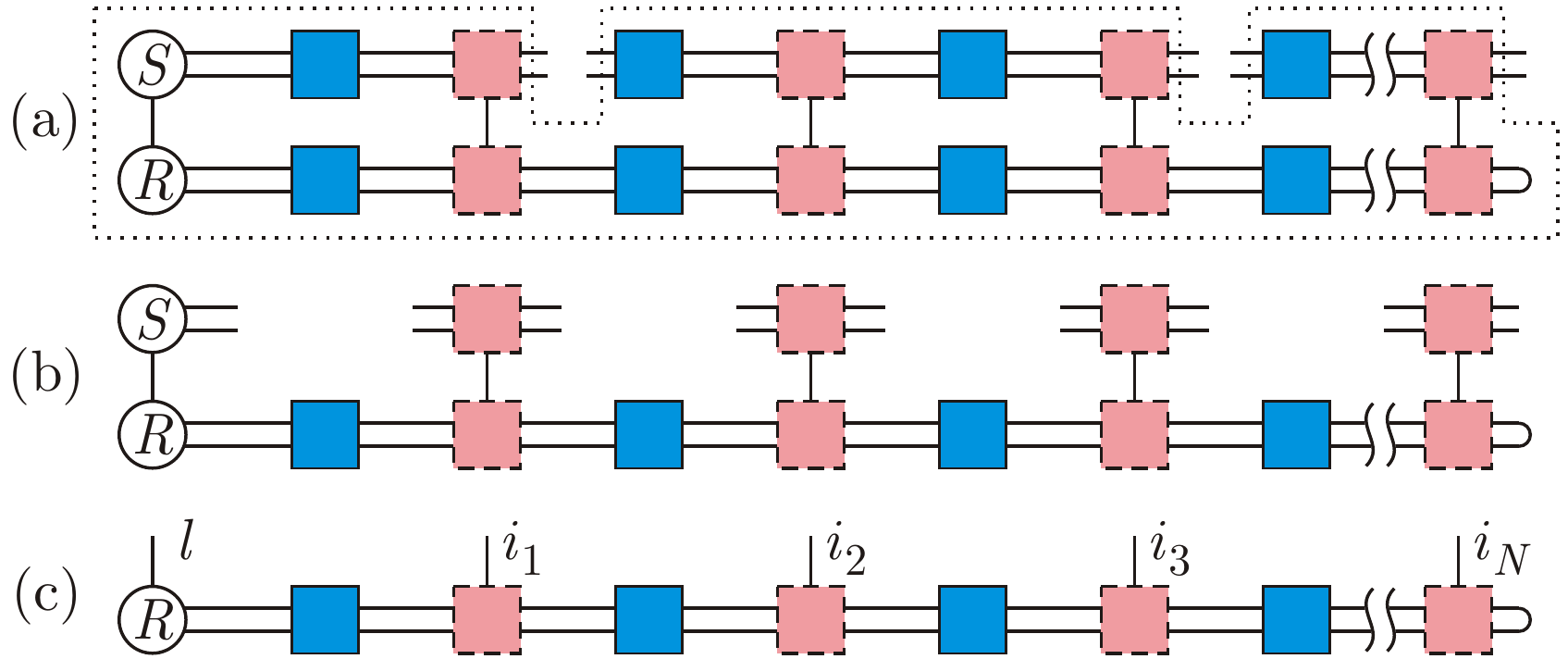}
\caption{(a) Process tensor ${\cal
T}(t_1;t_3;t_n)$~\cite{milz-2017,pollock-2018,modi-2018,milz-2018}.
(b) Influence
functional~\cite{makri-1995,sim-2001,strathearn-2017,strathearn-2018,pollock-2019}.
(c) Timeline reservoir network.} \label{figure3}
\end{figure}

A key object in our study is the tensor in Fig.~\ref{figure3}(c),
which we call a \textit{timeline reservoir network} (TRN). The TRN
contains all the information on the reservoir and controls all
features of open dynamics, including dissipation, Lamb shift, memory
effects like revivals~\cite{Breuer2002}. From the computational
viewpoint, for a fixed time $t$, TRN is a tensor with $t/{\tau}$
indices. Since the physical reservoir has a finite memory depth
$T$, the considered tensor must have vanishing correlations
between apart indices. Tensors of such a type can be effectively
approximated by one-dimensional tensor networks of matrix-product
(MP) form.

The TRN is closely related to the recent reformulations of the
Feynman-Vernon path
integrals~\cite{makri-1995,sim-2001,strathearn-2017,strathearn-2018,pollock-2019}
and the process
tensor~\cite{milz-2017,pollock-2018,modi-2018,milz-2018}, see
Figs.~\ref{figure3}(a)--\ref{figure3}(b). In fact, the influence functional in
Fig.~\ref{figure3}(b) can be explicitly calculated in the case of
a bosonic bath linearly coupled to the system~\cite{feynman-1963}
but it remains difficult to
contract with the system initial state and unitary evolution
tensors, so in Refs.~\cite{makri-1995,sim-2001,strathearn-2017}
the contraction calculation is approximated by fixing a finite memory
depth $K = T/{\tau}$. References~\cite{strathearn-2018,pollock-2019}
further use MP approximation of the influence functional (with
rank $\lambda_{\max}$ and accuracy $\lambda_{c}$), which allows to
deal with longer memory depths. Since $\lambda_{\max}$ is
$d_{ER}^2$ in our model, we actually estimate the complexity
($\lambda_{\max}$) of the algorithm in
Ref.~\cite{strathearn-2018}.

\paragraph*{MP approximation of the TRN.}
From a mathematical viewpoint, the constructed TRN can be treated as a pure multipartite quantum state $\ket{\psi}$, where summation index $i_m = 0, \ldots, 2n$ at time moment $t_m = m \tau$ plays the role of the physical index assigned to the $m$th particle:
\begin{equation}
\ket{\psi} = \sum_{l, i_1, i_2, \ldots, i_N} \psi_{l i_1 i_2
\ldots i_N} \ket{l} \otimes \ket{i_1} \otimes \ket{i_2} \otimes
\ldots \otimes \ket{i_N}.
\end{equation}

\noindent The only difference between the TRN and $\ket{\psi}$
is the normalization: TRN$\propto \ket{\psi}$, $\ip{\psi}{\psi} =
1$.

A multipartite quantum state $\ket{\psi}$ with a finite
correlation length $L$ can be effectively described via MP
approximation~\cite{Orus2014}. The benefit of such an
approximation is that it is able to reproduce spatial correlations
among particles within the characteristic length $L$. The
effectiveness of approximation means that the bond dimension $r$
of the ancillary space (rank of MP state) is rather small.
Similarly, the TRN is able to effectively reproduce temporal
correlations within the period $T$ with a rather small dimension
$d_{ER}$ of effective reservoir. Since we deal with matrices,
$d_{ER}^2$ is equivalent to the rank of corresponding MP state.
The dimension of the physical space in the MP state is $2n+1$, where
$n$ is the number of the subsystem's degrees of freedom involved in
the interaction with environment. Note that $n \leq d_S^2$. The
physics of MP approximations for states and TRN is summarized in
Table~\ref{table1}.

\begingroup
\squeezetable
\begin{table}
\begin{ruledtabular}
\begin{tabular}{l|l}
MP approximation of states & MP approximation of TRN \\
\hline\hline
position of $m$'th particle in space & time moment $t_m = m \tau$ on timeline  \\
\hline
dimension of a particle's Hilbert & twice the number of subsystem's  \\
space & degrees of freedom plus one, $2n+1$  \\
\hline
rank, bond dimension $r$ & square of dimension    \\
(dimension of ancillary space) & of effective reservoir, $d_{ER}^2$  \\
\hline
correlation length, $L$ & reservoir correlation time, $T$  \\
\end{tabular}
\end{ruledtabular} \caption{\label{table1} Correspondence between physical descriptions of pure quantum states and TRN in MP approximation.}
\end{table}
\endgroup

Suppose $\ket{\psi^{(r)}}$ is a rank-$r$ MP approximation of the pure state $\ket{\psi}$. The approximation error $\epsilon(r)$ equals the Frobenius distance between $\ket{\psi^{(r)}}$ and $\ket{\psi}$. Consider $\ket{\psi}$ as a bipartite state, with parties being separated by a cut between the $m$th and $(m+1)$th particles. MP approximation effectively disregards low-weight contributions in the Schmidt decomposition of $\ket{\psi}$ with respect to such a cut. The approximation error $\epsilon(r)$ is related to the R\'{e}nyi entropy of order $\alpha$ ($0< \alpha <1$), $S_{\alpha}$ of a single reduced density operator as follows~\cite{Verstraete2006}:
\begin{equation}
\ln[\epsilon(r)] \leq \frac{1-\alpha}{\alpha} \left[ S_\alpha -
\ln\left(\frac{r}{1-\alpha}\right) \right],
\label{epsilon}
\end{equation}

\noindent from which one readily obtains the sufficient bond dimension guaranteeing the arbitrary desired accuracy $\epsilon$:
\begin{equation} \label{rank-sufficient}
r_{\rm suff}(\epsilon) = \min_{0 < \alpha < 1} (1-\alpha) \epsilon^{-\alpha/(1-\alpha)} \, \exp(S_\alpha).
\end{equation}

\begin{figure}
\includegraphics[width=8cm]{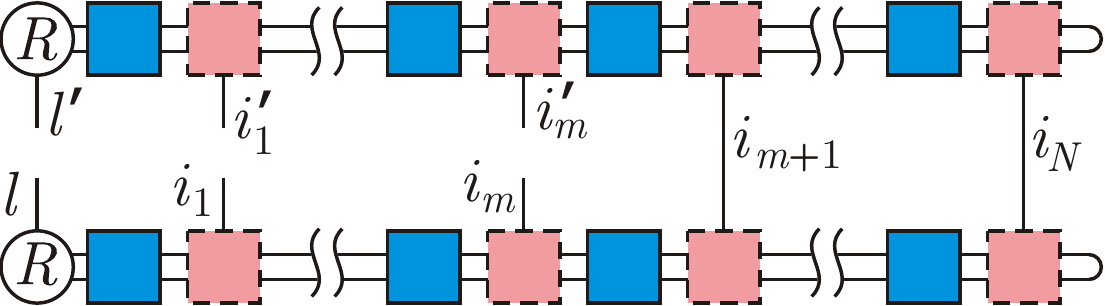}
\caption{Reduced matrix of TRN.} \label{figure4}
\end{figure}

In the language of TRN, the sufficient rank is the square of minimal dimension of effective reservoir, $d_{ER}^2$, which can reproduce all temporal correlations with error $\epsilon$. Therefore, to find $d_{ER}$ one needs to estimate the R\'{e}nyi entropy $S_{\alpha}$ of tensor in Fig.~\ref{figure4} considered as a matrix $M$ with 2 multi-indices $(l,i_1,\ldots,i_m)$ and $(l',i_1',\ldots,i_m')$:
\begin{eqnarray}\label{M}
&& M_{(l,i_1,\ldots,i_m),(l',i_1',\ldots,i_m')} \nonumber\\
&& = \sum_{i_{m+1},\ldots,i_N}
\psi_{l,i_1,\ldots,i_m,i_{m+1},\ldots,i_N}
\psi^{\ast}_{l',i_1',\ldots,i_m',i_{m+1},\ldots,i_N}. \qquad
\end{eqnarray}

The R\'{e}nyi entropy reads \cite{supmat}: 
\begin{equation} \label{S-from-above-final}
S_{\alpha} \lesssim \frac{1}{1-\alpha} \ln \frac{\left[
1+2n(\gamma\tau)^{\alpha}\right]^{T/\tau}}{(1+2n\gamma\tau)^{\alpha
T /\tau}}\approx 2n \gamma T \, \frac{(\gamma
\tau)^{\alpha-1}-\alpha}{1-\alpha} .
\end{equation}

\noindent The entropy $S_{\alpha}$ is a measure of the time
correlations in TRN. Substituting~\eqref{S-from-above-final}
in~\eqref{rank-sufficient}, we obtain the sufficient rank $r_{\rm
suff}$ of MP approximation of TRN with desired physical properties
(parameters $\gamma$, $n$, $T$, $\tau$) and accuracy $\epsilon$.
On the other hand, the rank of MP approximation is the square of
the dimension $d_{ER}$ of the \emph{effective}
reservoir that can reproduce all the features of open dynamics
(including memory effects) with accuracy $\epsilon$. Therefore, it
is possible to simulate the complex open system dynamics by using
the effective reservoir of dimension
\begin{equation} \label{der-sufficient}
d_{ER}(\epsilon) = \min_{0 < \alpha < 1} \frac{\sqrt{1-\alpha}}{\epsilon^{\alpha/2(1-\alpha)}} \exp \left[ n \gamma T \, \frac{(\gamma \tau)^{\alpha-1}-\alpha}{1-\alpha}  \right].
\end{equation}

Once the effective reservoir is constructed, the aggregate
``$S+ER$'' experiences the semigroup dynamics. This follows from
the tensor network representation in Fig.~\ref{figure2}. The TRN
has a homogeneous structure and so does its MP approximation. The
regular structure of building blocks in the time scale means that
the same transformation ${\mathds 1} + \tau {\cal L}$ acts on
``$S+ER$'' between the successive times $m\tau$ and $(m+1)\tau$.
The GKSL generator ${\cal L}$~\cite{gks-1976,lindblad-1976} acts
on $d_S d_{ER} \times d_S d_{ER}$ density matrices and guarantees
complete positivity of evolution.

\begin{figure}
    \includegraphics[width=8cm]{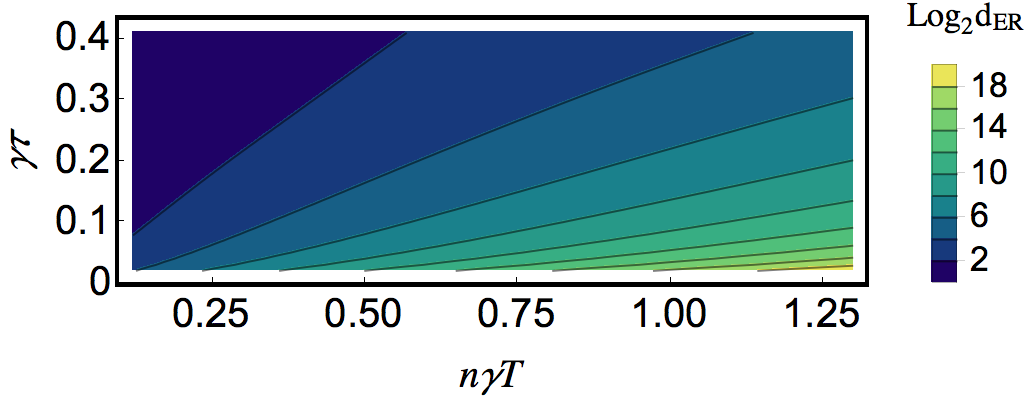}
    \caption{Number of qubits $\log_2(d_{ER})$ in effective reservoir, which is sufficient for simulation of open dynamics with accuracy $\epsilon = 0.05$. $n\gamma T$ is dimensionless memory time and $\gamma\tau$ is dimensionless minimal timescale.}
    \label{figure5}
\end{figure}

\paragraph*{Discussion.} Although the actual environment consists of infinitely many modes,
the developed theory facilitates the simulation of complex open
system dynamics with a finite dimensional effective reservoir. The
sufficient dimension $d_{ER}$ depends on two combinations of
physical parameters: $n\gamma T$ and $\gamma\tau$.
Figure~\ref{figure5} shows that one can simulate the open dynamics
on a classical computer for a wide range of parameters: $n\gamma T$
and $\gamma\tau$, accounting for all potential initial
correlations between the system and its environment.

\begin{figure}[b]
    \centering
    \includegraphics[width=8.5cm]{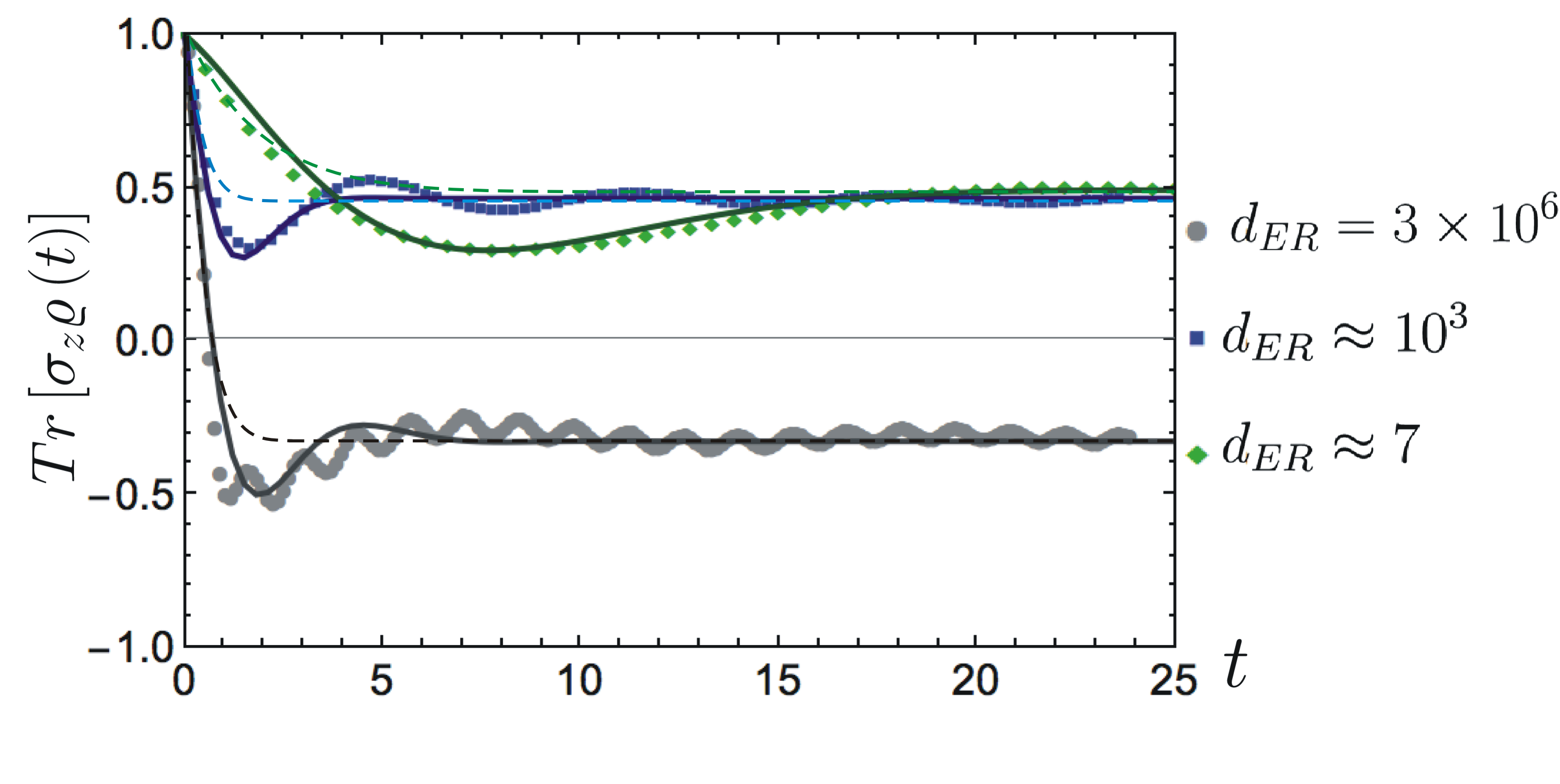}
    \caption{Typical evolutions of parameter ${\rm tr}[\sigma_z \rho(t)]$ of open qubit system $\rho(t)$ for different values of simulation complexity $d_{ER}(0.05)$ \cite{supmat}. Dotted lines depict the exact dynamics. Dashed lines depict the best Markov approximations ($d_{ER}' = 1$). Solid lines are the approximations obtained with the reservoir of fixed dimension $d_{ER}'' = 2$.}
    \label{figure6}
\end{figure}

The first illustrative example is a decay of the two-level system
(qubit) in multimode environment \cite{supmat}. We compare
the exactly solvable qubit dynamics, the Markov approximation
($d_{ER}'=1$), and the approximation obtained with the reservoir
of fixed dimension ($d_{ER}''=2$). Figure~\ref{figure6} shows that
the best Markovian approximation cannot reproduce oscillations in
the exact dynamics, whereas the approximation with the fixed
dimension $d_{ER}''=2$ fits well the exact dynamics when the
simulation complexity $d_{ER} \sim d_{ER}''$. However, if $d_{ER}$
is several orders of magnitude larger than $d_{ER}''$, then the
approximation is not able to reproduce memory effects present in
the exact solution. Thus, $d_{ER}$ does quantify the complexity of
dynamics.

The second example is the double quantum dot charge qubit coupled
to piezoelectric acoustic phonons~\cite{eckel-2006}. Here, $n=1$,
$T \approx 4 \tau$, $\tau = \omega_c^{-1}$, $\gamma =
0.05\omega_c$, and $\omega_c = 83$~GHz is the cutoff frequency of
the spectral function. Eq.~\eqref{der-sufficient} yields ${\rm
log}_2 d_{ER} = 4$ for $\epsilon=0.05$, i.e. the non-Markovian
qubit dynamics can be embedded into a Markovian evolution of the
very qubit and 4 auxiliary ones.

The third example is the non-Markovian evolution of the qubit due
to interaction with the dissipative
pseudomode~\cite{garraway-1997,mazzola-2009}: $\frac{d\rho}{dt}
= -i[H_0,\rho]+ \Gamma (a \rho a^{\dag} - \frac{1}{2}
\{a^{\dag}a,\rho\})$, where $H_0 = \omega_0 \sigma_+\sigma_- +
\omega a^{\dag}a + \Omega_0\sigma_x (a^{\dag} + a)$ and $\rho$
is the density operator for the qubit and the pseudomode. Physical
parameters are $n=1$, $T=\Gamma^{-1}$, $\tau = \omega^{-1}$, $\gamma =
\Omega_0 \sqrt{n_0+1}$, where $n_0$ is the effective number of
photons in the pseudomode. Our result, Eq.~\eqref{der-sufficient},
estimates where the pseudomode oscillator can be truncated (Fock
states with number of photons less than $d_{ER}$) to reproduce the
system dynamics with precision $\epsilon$ at any time despite the
memory effects and the counter-rotating terms in $H_0$. In this
example, construction of the effective reservoir reduces to the
subspace spanned by $d_{ER}$ lowest energy states of the
pseudomode because the particular dissipator forces
the pseudomode to the ground state.

There are physical scenarios in which the structure of the
effective reservoir follows from the model. For instance, in a
nitrogen-vacancy center in diamond, the inherent nitrogen
($^{14}$N) nuclear spin $(I=1)$ serves as an effective reservoir
for the electronic spin qubit~\cite{haase-2018}. In this case,
$d_{ER}=3$. Similarly, in a composite bipartite collision
model~\cite{lorenzo-2017}, $d_{ER}$ is given by the size of an
ancillary system. In general, however, the structure of the
effective reservoir is to be determined from the experimental
data. Reference~\cite{luchnikov-ml-2019} proposes a machine learning
algorithm to reconstruct the generator ${\cal L}$ based on a
series of repeated measurements on the open system.

Finally, our result is applicable to the influence functional
tensor networks in Refs.~\cite{strathearn-2018,pollock-2019},
where the analytical solution for open dynamics is not accessible,
and provides the upper bound on the maximum bond dimension,
$\lambda_{\max} < d_{ER}^2$. Conversely, for a fixed
computationally tractable size of bond dimension, e.g.,
$\lambda_{\max} \sim 10^3$, our result provides the region of
physical parameters $\gamma \tau$ and $n \gamma T$, for which the
algorithm in Ref.~\cite{strathearn-2018} definitely works well.

Importantly, the TRN is a multidimensional tensor, so it can be
approximated with MP form but also with other
constructions like multiscale entanglement renormalization
ansatz~\cite{MERA,Evenbly2014} or artificial neural
networks~\cite{Carleo2017,Amin2018}. The benefit of such networks
is that time correlations in the environment do not have to decay
exponentially as for the MP approximation.

\paragraph{Conclusion.} We gave a definition of simulation complexity of open quantum
dynamics in terms of a reservoir's effective dimension. We showed
that the tensor networks approach can be utilized to analyze
memory effects in open dynamics. We provided an estimation of
simulation complexity using a set of physical parameters. Our
estimation is universal and fits well the arbitrary open
quantum dynamics with finite memory.

\paragraph*{Acknowledgments}
The authors thank Alexey Akimov, Mikhail Krechetov, Eugene
Polyakov, Alexey Rubtsov, and Mario Ziman for fruitful
discussions. The study is supported by Russian Foundation for
Basic Research under Project No. 18-37-20073. I.A.L. thanks Russian
Foundation for Basic Research for partial support under Project
No. 18-37-00282.

\begin{widetext}

\section*{Supplemental Material}

\subsection{Upper bound on R\'enyi entropy}
Here we discuss the upper bound on R\'enyi entropy to clarify the
derivation of formula~(12) in the main text. Consider an arbitrary
density matrix $M$ that has the form $M = \sum_q {\bf v}_q {\bf
v}_q^{\dag}$, where the set of vectors $\{{\bf v}_q \}_q$ is not
orthogonal nor normalized. Let us prove that the R\'enyi entropy
$S_\alpha (M) = \frac{1}{1-\alpha} \ln {\rm tr}M^\alpha$ satisfies
the inequality $S_\alpha (M) \leq \frac{1}{1-\alpha}\ln \sum_q
\left( {\bf v}_q^{\dag} {\bf v}_q \right)^{\alpha}$ if
$0<\alpha<1$.

In fact, the classical R\'enyi entropy $H_\alpha$ is Schur-concave
for all $\alpha > 0$ (see, e.g., \cite{sason-2018}), i.e.
$H_{\alpha}(R) \leq H_{\alpha}(P)$ whenever the probability
distribution $P$ is majorized by the probability distribution $R$
($\sum_{i=1}^k P_i^{\downarrow} \leq \sum_{i=1}^k
R_i^{\downarrow}$ for all $k=1,2, \ldots$). By Theorem~10 of
Ref.~\cite{nielsen-2001} the equality $M = \sum_q {\bf v}_q {\bf
v}_q^{\dag}$ implies that the distribution $V = \{ {\bf
v}_q^{\dag}{\bf v}_q \}_q$ is majorized by the vector $\Lambda$ of
eigenvalues of $M$. Therefore,
\begin{eqnarray}
S_\alpha (M) = H_{\alpha}(\Lambda) \leq H_{\alpha}(V) =
\frac{1}{1-\alpha}\ln \sum_q \left( {\bf v}_q^{\dag} {\bf v}_q
\right)^{\alpha}.
\end{eqnarray}

\subsection{Diagrammatic language for tensor networks}
In this section we demonstrate key physical objects using tensor network representation (language).

Fig.~\ref{TRN_sequence}(a) illustrates the timeline reservoir network (TRN) as an analogue of multipartite quantum state $\ket{\psi}$ . It also depicts the TRN matrix in analogy with the density matrix $\ket{\psi}\bra{\psi}$ for states.

In Fig.~\ref{TRN_sequence}(b) we simplify the notation and coarse grain the elements of the tensor network. We replace double lines by single thick lines. The tracing element in the right hand side of network is depicted as a semicircle.

Fig.~\ref{TRN_sequence}(c) depicts TRN matrix in terms of new notation.

In Fig.~\ref{TRN_sequence}(d) we depict the reduced TRN matrix, which is constructed in analogy with the reduced density matrix ($\rho_{s} ={\rm tr}_{r}{\rho_{s+r}}$). Connected indices between upper and lower parts of network mean summation, i.e. reduction over non-relevant part with time $t \geq m\tau$.

\begin{figure}
    \centering
    \includegraphics[scale = 1.2]{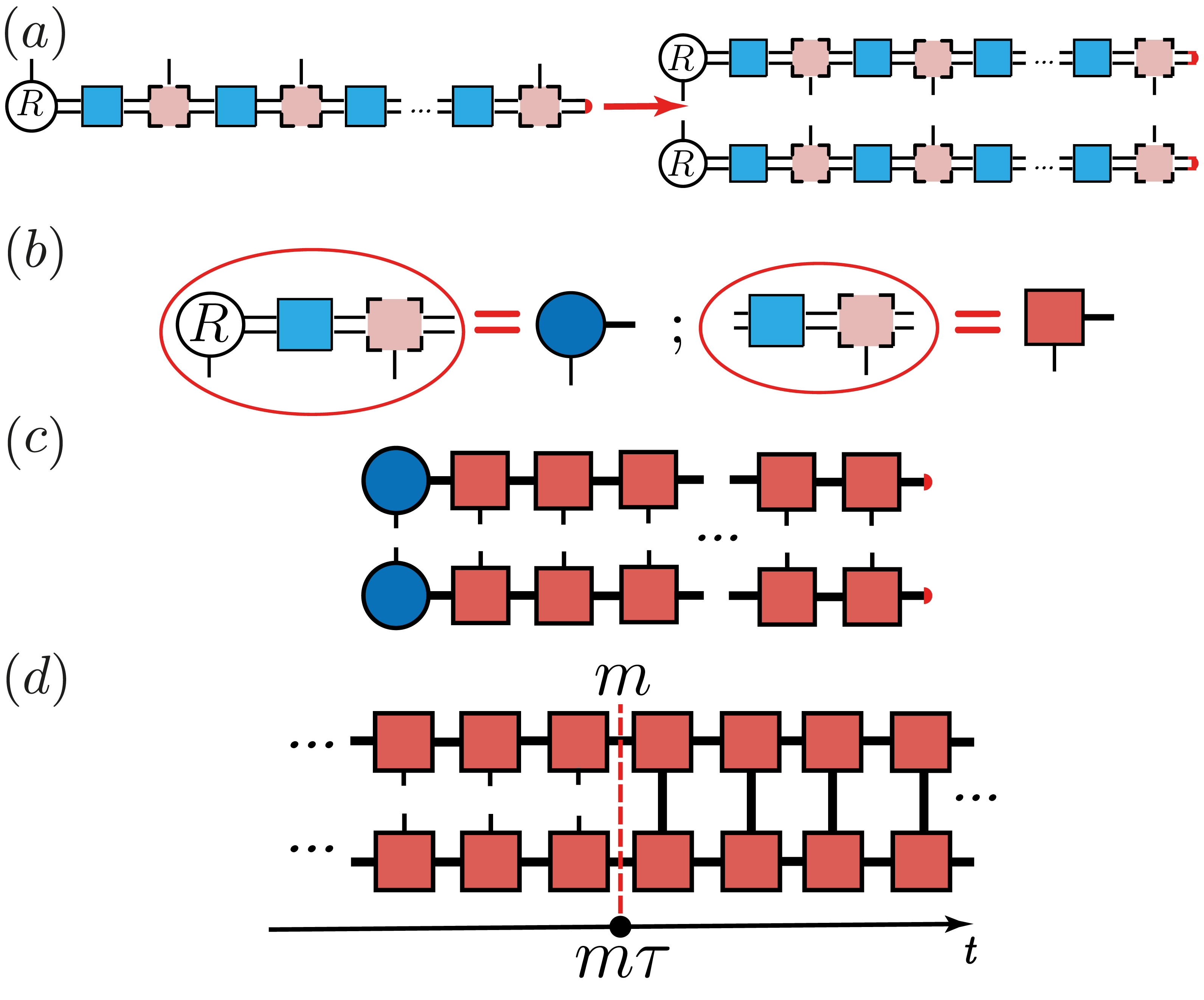}
    \caption{\label{TRN_sequence} (a) TRN and TRN matrix. (b) Change of notation to simplify tensor network representation.
    (c) TRN matrix in simplified notation.
    (d) Reduced matrix of TRN.}

    \includegraphics[scale = 1.2]{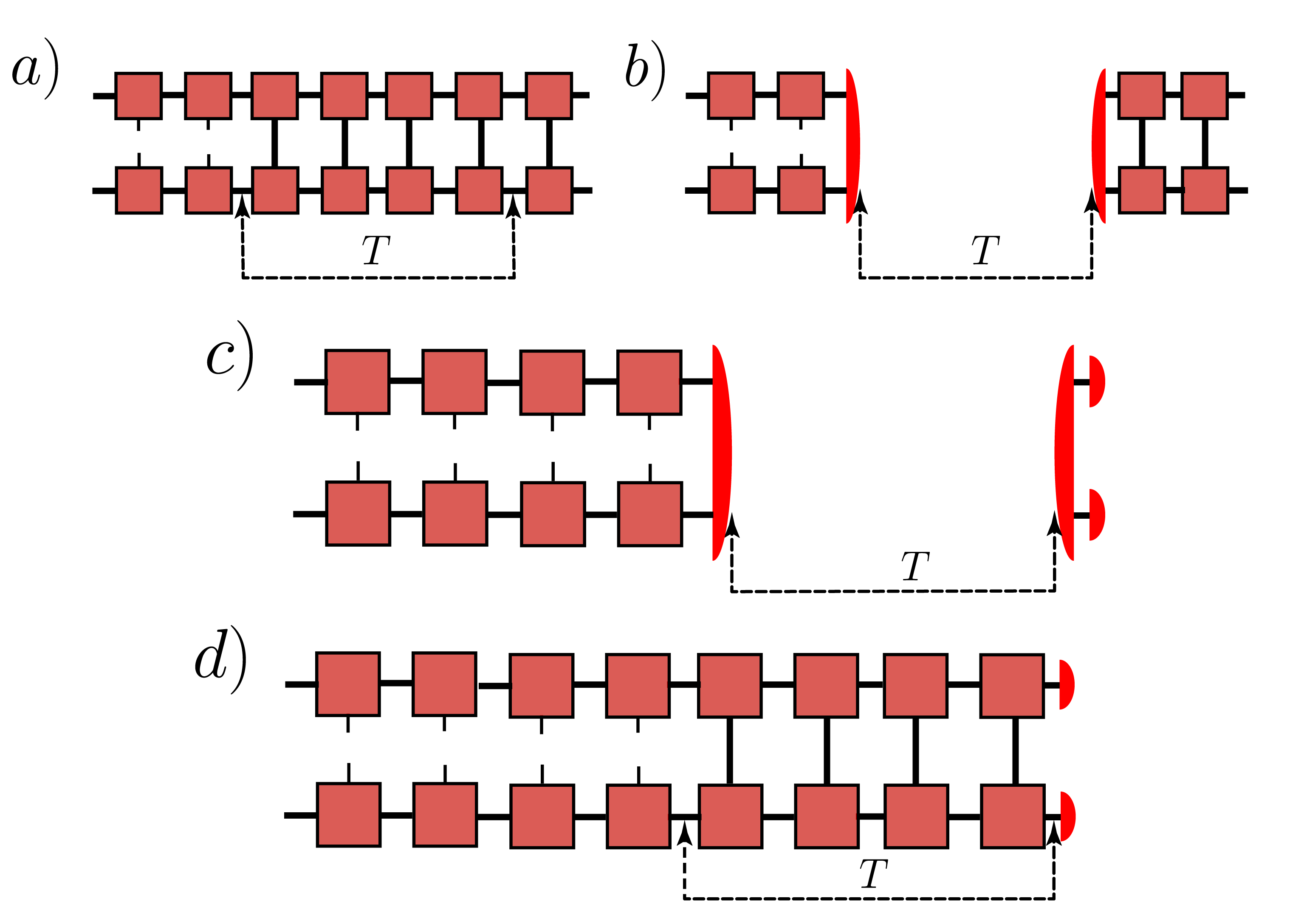}
    \caption{\label{truncation} (a) Reduced matrix of TRN and the reservoir correlation time $T$. (b) The absence of time correlations between blocks separated by time $T$. (c) The right hand side is a rank-0 tensor. (d) The marked tensor of length $T$ is exactly the element $=\!\!\!|\!\!\!\supset$ in Fig.~\ref{truncation}(c).}
\end{figure}

\begin{figure}
    \centering
    \includegraphics[scale = 1.5]{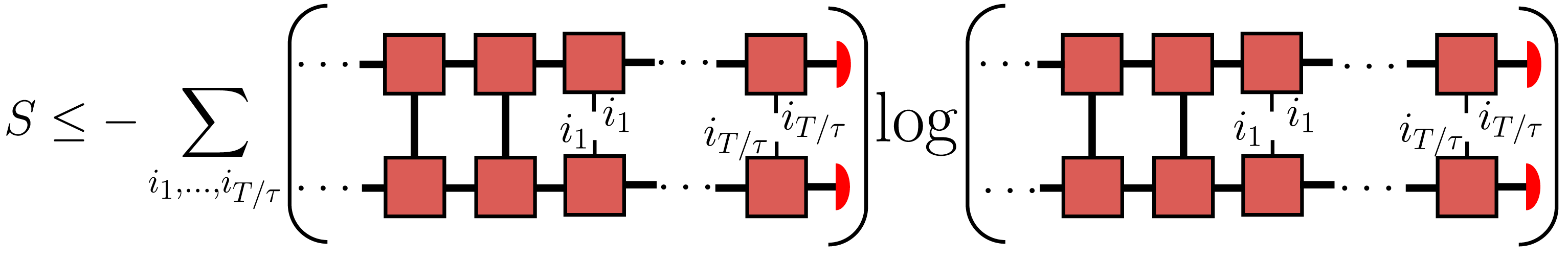}
    \caption{Estimation of the von Neumann entropy of the reduced TRN within the framework of diagrammatic representation.}
    \label{Entropy}
\end{figure}

Given a reduced matrix $M$ of TRN, namely,
\begin{equation}
M_{(l,i_1,\ldots,i_m),(l',i_1',\ldots,i_m')} =
\sum_{i_{m+1},\ldots,i_N}
\psi_{l,i_1,\ldots,i_m,i_{m+1},\ldots,i_N}
\psi^{\ast}_{l',i_1',\ldots,i_m',i_{m+1},\ldots,i_N}.
\end{equation}

\noindent one needs to estimate its R\'enyi entropy $S_{\alpha}$.
The result of section A is that $S_{\alpha}(M) \leq
(1-\alpha)^{-1} \ln \sum_q \left( {\bf v}_q^{\dag} {\bf v}_q
\right)^{\alpha}$. In our case $q= (i_{m+1},\ldots,i_N)$ and ${\bf
v}_q$ is a vector with components $\psi_{l,i_1,\ldots,i_m,q}$, so
we have
\begin{equation} \label{S-from-above}
S_{\alpha} \leq (1-\alpha)^{-1} \ln \!\!\!\!\!
\sum_{i_{m+1},\ldots,i_N} \left( \sum_{l,i_1,\ldots,i_m} \!\!
\vert \psi_{l,i_1,\ldots,i_m,i_{m+1},\ldots,i_N} \vert^2
\right)^{\!\!\alpha}.
\end{equation}

Let us consider the physical structure of $M$, see
Fig.~\ref{truncation}. In Fig.~\ref{truncation}(a), we explicitly
mark out the reservoir correlation time $T$. Since temporal
correlations decay within the time period $T$, we assume that time
correlations between nodes separated by time $t \ge T$ can be
neglected. This means that the action of tensor network of length
$T$ can be replaced by a simpler tensor of the form
$=\!\!\!|\!\!\!\supset \ \subset\!\!\!|\!\!\!=$, see
Fig.~\ref{truncation}(b). Thus,
\begin{equation}
=\!\!\!(~~\text{tensor of length  } t \ge T~~)\!\!\!=  \ \ \ \ \  \Longleftrightarrow \ \ \ \ \ =\!\!\!|\!\!\!\supset \ \ \subset\!\!\!|\!\!\!= \ .
\end{equation}

Concatenation of the tensor $\subset\!\!\!|\!\!\!=$ with the rest
of right-hand side tensor is merely a number (rank-0 tensor),
Fig.~\ref{truncation}(c). In other words, the contribution of
indexes $i_{m+T/\tau+1},i_{m+T/\tau+2},\ldots,i_{N}$ in formula
(\ref{S-from-above}) is trivial: the summation over indexes
$i_{m+T/\tau+1},i_{m+T/\tau+2},\ldots,i_{N}$ does not change the
structure of the matrix $M$ as it only leads to a multiplication
factor. The reduced matrix $M$ of TRN is not affected by time
moments happening after the reservoir correlation timescale has
passed.

That observation explains the fact why in
formula~(\ref{S-from-above}) one can replace the tensor
$\psi_{l,i_1,\ldots,i_m,i_{m+1},\ldots,i_N}$ by a tensor
$\widetilde{\psi}_{l,i_1,\ldots,i_m,i_{m+1},\ldots,i_{m+T/\tau}}$
of finite length and appropriate normalization
$\sum_{l,i_1,\ldots,i_{m+T/\tau}}
\vert\widetilde{\psi}_{l,i_1,\ldots,i_{m+T/\tau}} \vert^2 = 1$.

Therefore, the R\'enyi entropy of reduced TRN matrix equals the
R\'enyi entropy of a tensor depicted in the left-hand side of
Fig.~\ref{truncation}(c). The element $=\!\!\!|\!\!\!\supset$ in
the left-hand side of Fig.~\ref{truncation}(c) can be replaced by
a tensor of finite length $T$, and we end up with a tensor
network illustrated in Fig.~\ref{truncation}(d). Following the
notation of the main text,
\begin{equation}
\text{Fig.~\ref{truncation}(d)}   \ \ \ \Longleftrightarrow \ \ \ \sum_{i_{m+1},\ldots,i_{m+T/\tau}}
\widetilde{\psi}_{l,i_1,\ldots,i_m,i_{m+1},\ldots,i_{m+T/\tau}}
\widetilde{\psi}^{\ast}_{l',i_1',\ldots,i_m',i_{m+1},\ldots,i_{m+T/\tau}} \, .
\end{equation}

\noindent and we get
\begin{equation} \label{S-from-above-trun}
S_{\alpha} \lesssim (1-\alpha)^{-1} \ln \!\!\!\!\!
\sum_{i_{m+1},\ldots,i_{m+T/\tau}} \left( \sum_{l,i_1,\ldots,i_m}
\!\! \vert
\widetilde{\psi}_{l,i_1,\ldots,i_m,i_{m+1},\ldots,i_{m+T/\tau}}
\vert^2 \right)^{\!\!\alpha}.
\end{equation}

Recalling the explicit structure of TRN given by formulas (5)--(7)
in the main text, we find
\begin{equation}
\sum_{l,i_1,\ldots,i_m} \vert
\widetilde{\psi}_{l,i_1,\ldots,i_m,i_{m+1},\ldots,i_{m+T/\tau}}
\vert^2 \approx (1+2n\gamma\tau)^{-T/\tau} \prod_{p=1}^{T/\tau}
\left\{
\begin{array}{ll}
  1, &  i_{m+p} = 0,  \\
  \gamma\tau, &  i_{m+p} = 1,\ldots,2n ~. \\
\end{array}
\right.
\end{equation}

Substituting this result in (\ref{S-from-above-trun}), we obtain
\begin{equation} %\label{S-from-above-final}
S_{\alpha} \lesssim \frac{1}{1-\alpha} \ln \frac{\left[
1+2n(\gamma\tau)^{\alpha}\right]^{T/\tau}}{(1+2n\gamma\tau)^{\alpha
T /\tau}}\approx 2n \gamma T \, \frac{(\gamma
\tau)^{\alpha-1}-\alpha}{1-\alpha} .
\end{equation}

To finalize the explanation of diagrammatic language in the
estimation of entropy $S_{\alpha}$, in Fig.~\ref{Entropy} we
depict the case $\alpha \rightarrow 1$ when $S_{\alpha}$ tends to
the von Neumann entropy.

\subsection{Exactly solvable model}
In this section, we consider a particular example of an exactly solvable open quantum dynamics. Let $S$ be a two-level system with energy separation $\Omega$ and $R$ be a bosonic bath at zero temperature. The total Hamiltonian reads (hereafter the Planck constant $\hbar = 1$)
\begin{eqnarray}\label{Hamiltonian}
H=\Omega\sigma_{+}\sigma_{-}+\sum_{m}\omega_{m}b_{m}^{\dagger}b_{m} +\sum_{m}g_{m}\left(b_{m}^{\dagger}\sigma_{-} + b_{m}\sigma_{+}\right),
\end{eqnarray}

\noindent where $\sigma_+ = \ket{1}\bra{0}$ and $\sigma_- = \ket{0}\bra{1}$, $\ket{0}$ and $\ket{1}$ are the ground and excited states of the two-level system, respectively, index $m$ enumerates bosonic modes, $b_m$ and $b_m^{\dagger}$ are the annihilation and creation operators for bosons in $m$'th mode, $g_m$ is the coupling strength given by formula
\begin{eqnarray}
\label{coupling}
g_{m}=
\left\{
    \begin{array}{ll}
        g & \text{if~~} \omega_{\rm min}\leq\omega_{m}\leq\omega_{\rm max}, \\
        0 & \text{otherwise}.
    \end{array}
\right.
\end{eqnarray}

Consider the equidistant set of mode frequencies $\omega_m$ with $\omega_{m+1}-\omega_m = \Delta\omega$, then the bath consists of $N = \frac{\omega_{\rm{max}} - \omega{\rm{min}}}{\Delta\omega}$ modes with freqencies $\omega_m = \omega_{\rm min} + m\Delta\omega$, $m=1,\ldots,N$.

The initial state of ``$S$+$R$'' is pure and has the form
\begin{eqnarray}
\ket{\psi(0)}=\ket{1,\rm vac},
\end{eqnarray}

\noindent which implies that the two-level system is in the excited state and that the bath has no excitation. The Hamiltonian~(\ref{Hamiltonian}) preserves the number of excitations, so the solution of the Schr\"{o}dinger equation $i\frac{d}{dt}\ket{\psi(t)} = H \ket{\psi(t)}$ has the form
\begin{eqnarray}\label{ANZATS}
\ket{\psi(t)}=\alpha(t)\ket{1,{\rm vac}}+\sum_m\beta_m(t)\ket{0,1_m},
\end{eqnarray}

\noindent where the coefficients $\alpha(t)$ and $\beta_m(t)$ satisfy the following system of differential equations
\small{
\begin{eqnarray}
i\frac{d}{dt}\begin{bmatrix}
    \alpha(t)\\
    \beta_1(t)\\
    \beta_2(t)\\
    \vdots\\
    \beta_{N}(t)
    \end{bmatrix}=\begin{bmatrix}
    \Omega&&g&&g&&\dots&&g\\
    g&&\Delta\omega&&0&&\dots&&0\\
    g&&0&&2\Delta\omega&&\dots&&0\\
    \vdots&&\vdots&&\vdots&&\ddots&&\vdots\\
    g&&0&&0&&\dots&&N\Delta\omega
    \end{bmatrix}\begin{bmatrix}
    \alpha(t)\\
    \beta_1(t)\\
    \beta_2(t)\\
    \vdots\\
    \beta_{N}(t)
    \end{bmatrix}
\end{eqnarray}}

\noindent and meet the initial requirements $\alpha(0)=1$ and $\beta_m(0)=0$.

The parameter $\alpha(t)$ is a solution of:
\begin{equation}
\frac{d}{dt}\alpha(t) = -i\Omega\alpha(t) - \sum_{k=1}^N g^2 \int_0^t \exp\left[-ik\Delta\omega(t-t')\right]\alpha(t') {\rm d}t'.
\end{equation}

Assume $\Delta\omega \rightarrow 0$, we replace the summation $\sum_k$ by the integration:
\begin{eqnarray}
\sum_{k=1}^{N} \rightarrow \frac{1}{\Delta \omega} \int_{\omega_{\rm min}}^{\omega_{\rm max}} {\rm d}\omega.
\end{eqnarray}

\noindent Therefore, we obtain the model with reservoir containing an infinite number of modes. Nonetheless, such a model is still solvable as the parameter $\alpha(t)$ is a solution of the following integrodifferential equation:
\begin{equation}
\frac{d}{dt}\alpha(t)=-i\Omega\alpha(t)-\frac{g^2}{\Delta \omega}\int_{0}^{t} G(t-t') \alpha(t') {\rm d}t', \qquad G(t-t') = \int_{\omega_{\rm min}}^{\omega_{\rm max}} \exp\left[-i\omega(t-t')\right] {\rm d}\omega.
\end{equation}

Formally $\alpha(t)$ reads
\begin{eqnarray}
\alpha(t)=\frac{1}{2\pi i}\int_{-\infty+\varepsilon}^{\infty+\varepsilon}{\rm d}s e^{st}\frac{1}{s+i\Omega+\frac{g^2}{\Delta\omega}\int_{\omega_{\rm max}}^{\omega_{\rm max}}{\rm d}\omega\frac{1}{s+i\omega}},
\end{eqnarray}

\noindent and the quantity ${\rm tr}[\sigma_z \rho(t)]$ illustrated in Fig.~6 in the main text equals ${\rm tr}[\sigma_z \rho(t)] = 2|\alpha(t)|^2 -1$.

Now, let us define the parameters $T$, $\tau$, $\gamma$, $n$ needed for the estimation of simulation complexity (the dimension of effective reservoir, $d_{ER}$). The memory kernel $G$ decays significantly within the reservoir correlation time $T$, which in our case equals $T=\frac{1}{\omega_{\rm max}-\omega_{\rm min}}$. The minimal timescale of evolution is $\tau=\frac{1}{\omega_{\rm max}}$. Comparing the general form of interaction Hamiltonian $H_{\rm int} = \gamma\sum_{i=1}^{n}A_i\otimes B_i$ and the interaction Hamiltonian in our example $g\sum_{m=1}^{\frac{\omega_{\rm max}-\omega_{\rm min}}{\Delta\omega}}(b^\dagger_m\sigma_-+b_m\sigma_+)$, we conclude that $n=2$ and $\gamma = g\frac{\omega_{\rm max}-\omega_{\rm min}}{\Delta\omega}$. Fixing the accuracy $\epsilon=0.05$ and substituting parameters $T$, $\tau$, $\gamma$, $n$ in formula (15) in the main text, we calculate the sufficient dimension of effective reservoir $d_{ER}$ for three different evolutions depicted in Fig.~6 in the main text.

\subsection{Semigroup evolution of system ($d_{ER}'=1$)}

Direct application of the weak-coupling and Born--Markov
approximations~\cite{Breuer2002} to the model in section C results
in the exponential decay ${\rm tr}[\sigma_+ \sigma_- \rho(t)] =
\exp(-\Gamma t)$ because the reservoir is assumed to be
time-independent (always in a vacuum state). Such a behavior is in
strong contrast to the exact solution in Fig.~6 in the main text,
where the steady state has a non-zero population ${\rm
tr}[\sigma_+ \sigma_- \rho(\infty)]$.

A general semigroup evolution of the very qubit corresponds to the
situation, when one forcibly fixes the dimension $d_{ER}'=1$.

Let us consider the semigroup evolution of the qubit with a GKSL
generator, which takes into account the finite population of
excited state:
\begin{equation}
\frac{d\rho}{dt} = -i \Omega \left[\sigma_+\sigma_-,\rho\right]
+ \gamma_{\downarrow} \left(\sigma_- \rho \sigma_+ -
\frac{1}{2}\{\sigma_+\sigma_-,\rho\}\right) + \gamma_{\uparrow}
\left(\sigma_+ \rho \sigma_- -
\frac{1}{2}\{\sigma_-\sigma_+,\rho\}\right).
\end{equation}

Note that this equation is nothing else but the generalized
amplitude damping process~\cite{nielsen-2000}. Adjusting the
relaxation rates $\gamma_{\downarrow}$ and $\gamma_{\uparrow}$, we
get the best Markov approximations shown in Fig.~6 in the main
text. This figure also illustrates that the best Markov
approximation cannot reproduce non-Markovian memory effects, the
correct effective dimension $d_{ER}$ being much greater than
$d_{ER}'=1$.

\subsection{Semigroup evolution of system and reservoir of fixed dimension $d_{ER}''=2$}

Let us consider a model with the small reservoir of forcibly fixed
dimension $d_{ER}'' = 2$ such that the system and the small
reservoir altogether experience the semigroup dynamics. Such a
model would be a good approximation of the exact dynamics found in
section C if the simulation complexity $d_{ER} \sim d_{ER}''$.

We consider the following parameterization of the GKSL generator
${\cal L}$ governing the evolution equation $\frac{d\rho}{dt} = {\cal L}\rho$ for the density operator
$\rho(t)$ of the two level system and small reservoir:
\begin{eqnarray}
{\cal L}\rho(t) &=&-i\left[h,\rho(t)\right]+{\cal D}(\rho(t)),\nonumber\\
h&=&\Omega_1 \sigma_+\sigma_-\otimes {\mathds 1}+\Omega_2{\mathds 1}\otimes\sigma_+\sigma_- + \widetilde{g}\left(\sigma_+\otimes\sigma_-+\sigma_-\otimes\sigma_+\right),\nonumber\\
{\cal D}(\rho(t))&=&\gamma_{1,\downarrow}\left(\sigma_-\otimes {\mathds 1}\rho(t)\sigma_+\otimes{\mathds 1}-\frac{1}{2}\{\sigma_+\sigma_-\otimes {\mathds 1},\rho(t)\}\right)\nonumber\\
&&+\gamma_{1,\uparrow}\left(\sigma_+\otimes {\mathds
1}\rho(t)\sigma_-\otimes{\mathds
1}-\frac{1}{2}\{\sigma_-\sigma_+\otimes {\mathds
1},\rho(t)\}\right)\nonumber\\
&&+ \gamma_{2,\downarrow}\left({\mathds
1}\otimes\sigma_-\rho(t){\mathds
1}\otimes\sigma_+-\frac{1}{2}\{{\mathds
1}\otimes\sigma_+\sigma_-,\rho(t)\}\right)\nonumber\\
&& +\gamma_{2,\uparrow}\left({\mathds
1}\otimes\sigma_+\rho(t){\mathds
1}\otimes\sigma_--\frac{1}{2}\{{\mathds
1}\otimes\sigma_-\sigma_+,\rho(t)\}\right).
\end{eqnarray}

\noindent The parameter $\Omega_1$ is the energy difference
between excited and ground states of the system $S$ with a
potential contribution of the Lamb shift; $\Omega_2$ is the energy
separation between levels of the effective reservoir;
$\widetilde{g}$ is the coupling constant; $\gamma_{1,\downarrow}$
and $\gamma_{1,\uparrow}$ are decay rates of the system; and
$\gamma_{2,\downarrow}$ and $\gamma_{2,\uparrow}$ are decay rates
of the effective reservoir. All these parameters are adjusted to
approximate the exact dynamics described in section C.

Figure~6 in the main text shows that the simple model with
reservoir of fixed dimension $d_{ER}''=2$ adequately describes
non-Markovian dynamics with simulation complexity $d_{ER}(0.05)
\approx 7$. However, such a simple model fails to reproduce memory
effects associated with dynamics for which $d_{ER}(0.05) \sim
10^3$. This fact justifies that in order to reproduce memory
effects with simulation complexity $d_{ER}$ one has to deal with a
reservoir of comparable dimension.\\

\end{widetext}


\begin{thebibliography}{99}

\bibitem{Sutherland2004} B. Sutherland, \emph{Beautiful Models: 70 Years of Exactly Solved Quantum Many-Body Problems} (World Scientific, 2004).

\bibitem{Bethe1931} H. Bethe, Z. Phys. A \textbf{71}, 205 (1931).

\bibitem{Georges1996} A. Georges, G. Kotliar, W. Krauth, and M. J. Rozenberg, Rev. Mod. Phys. {\bf 68}, 13 (1996).

\bibitem{Kotliar2006} G. Kotliar, S. Y. Savrasov, K. Haule, V. S. Oudovenko, O. Parcollet, and C. A. Marianetti, Rev. Mod. Phys. {\bf 78}, 865 (2006).

\bibitem{Barnes1976} S. E. Barnes, J. Phys. F: Metal Phys. {\bf 6}, 1375 (1976).

\bibitem{Kotliar1986} G. Kotliar and A. E. Ruckenstein, Phys. Rev. Lett. {\bf 57}, 1362 (1986).

\bibitem{Andrei1980} N. Andrei, Phys. Rev. Lett. {\bf 45}, 379 (1980).

\bibitem{Wiegmann1980} P. B. Wiegmann, Phys. Lett. A {\bf 80}, 163 (1980).

\bibitem{Georges1992} A. Georges and G. Kotliar, Phys. Rev. B {\bf 45}, 6479 (1992).

\bibitem{Ouerdane2007} R. Fr\'esard, H. Ouerdane, and T. Kopp, Nucl. Phys. B {\bf 785}, 286 (2007).

\bibitem{Ouerdane2008} R. Fr\'esard, H. Ouerdane, and T. Kopp, EPL {\bf 82}, 31001 (2008).

\bibitem{Orus2014} R. Or\'us, Annals Phys. {\bf 349}, 117 (2014).

\bibitem{Orus2014b} R. Or\'us, Euro. Phys. J. B {\bf 87}, 280 (2014).

\bibitem{Verstraete2008} F. Verstraete, V. Murg, and J. I. Cirac, Advances Phys. {\bf 57}, 143 (2008).

\bibitem{White1992} S. R. White, Phys. Rev. Lett. {\bf 69}, 2863 (1992).

\bibitem{Schollwock2011} U. Schollw\"ock, Annals Phys. {\bf 326}, 96 (2011).

\bibitem{Schollwock2011b} U. Schollw\"ock, Phil. Trans. Royal Soc. London A {\bf 369}, 2643 (2011).

\bibitem{Landau2009} D. P. Landau and K. Binder, \emph{A guide to Monte Carlo simulations in statistical physics} (Cambridge University Press, Cambridge, 2009).

\bibitem{Loh1990} E. Y. Loh, Jr., J. E. Gubernatis, R. T. Scalettar, S. R. White, D. J. Scalapino, and R. L. Sugar, Phys. Rev. B {\bf 41}, 9301 (1990).

\bibitem{Lieb1972} E. Lieb and D. Robinson, Commun. Math. Phys. {\bf 28}, 251 (1972).

\bibitem{schollwock-2013} U. Schollw\"{o}ck, ``DMRG: Ground states, time evolution, and spectral functions,'' Chap. 16 in {\it Emergent Phenomena in Correlated Matter} edited by E. Pavarini, E. Koch, and U. Schollw\"{o}ck (Verlag des Forschungszentrum J\"{u}lich, 2013).

\bibitem{prosen-2007} T. Prosen and M. \v{Z}nidari\v{c}, Phys. Rev. E {\bf 75}, 015202(R) (2007).

\bibitem{InchWorm} H.-T. Chen, G. Cohen, and D. R. Reichman, J. Chem. Phys. {\bf 146}, 054106 (2017).

\bibitem{Davies1976} E. B. Davies, \emph{Quantum Theory of Open Systems}, (Academic Press, London, 1976).

\bibitem{Breuer2002} H.-P. Breuer and F. Petruccione, \emph{The Theory of Open Quantum Systems} (Oxford University Press, Oxford, 2002).

\bibitem{rau-1963} J. Rau, Phys. Rev. {\bf 129}, 1880 (1963).

\bibitem{palma-1996} G. M. Palma, K.-A. Suominen, and A. K. Ekert, Proc. R. Soc. London A {\bf 452}, 567 (1996).

\bibitem{fpmz-2017} S. N. Filippov, J. Piilo, S. Maniscalco, and M. Ziman, Phys. Rev. A {\bf 96}, 032111 (2017).

\bibitem{Alicki2012} R. Alicki and K. Lendi, \emph{Quantum Dynamical Semi-Groups and Applications}, Lecture Notes Physics {\bf 286} (Springer-Verlag, Berlin, 1987).

\bibitem{Holevo2012} A. S. Holevo, \emph{Quantum systems, channels, information. A mathematical introduction} (de Gruyter, Berlin/Boston, 2012).

\bibitem{piilo-2011} B.-H. Liu, L. Li, Y.-F. Huang, C.-F. Li, G.-C. Guo, E.-M. Laine, H.-P. Breuer, and J. Piilo, Nat. Phys. {\bf 7}, 931 (2011).

\bibitem{cirac-2011} C. Navarrete-Benlloch, I. de Vega, D. Porras, and J. I. Cirac, New J. Phys. {\bf 13},
023024 (2011).

\bibitem{ma-2012} J. Ma, Z. Sun, X. Wang, and F. Nori, Phys. Rev. A {\bf 85}, 062323
(2012).

\bibitem{hoope-2012} U. Hoeppe, C. Wolff, J. K\"{u}chenmeister, J. Niegemann, M. Drescher, H. Benner, and K. Busch, Phys. Rev. Lett. {\bf 108}, 043603 (2012).

\bibitem{yang-2013} W. L. Yang, J.-H. An, C. Zhang, M. Feng, and C. H.
Oh, Phys. Rev. A {\bf 87}, 022312 (2013).

\bibitem{hughes-2015} K. Roy-Choudhury and S. Hughes, Optica {\bf 2}, 434 (2015).

\bibitem{eisert-2015} S. Gr{\"o}blacher, A. Trubarov, N. Prigge, G. D. Cole, M. Aspelmeyer, and
J. Eisert, Nat. Commun. {\bf 6}, 7606 (2015).

\bibitem{cirac-2017} A. Gonz\'alez-Tudela and J. I. Cirac, Phys. Rev. Lett. {\bf 119}, 143602 (2017).

\bibitem{wittemer-2018} M. Wittemer, G. Clos, H.-P. Breuer, U. Warring, and T. Schaetz, Phys. Rev. A {\bf 97}, 020102(R)
(2018).

\bibitem{wang-2018} F. Wang, P.-Y. Hou, Y.-Y. Huang, W.-G. Zhang, X.-L. Ouyang,
X. Wang, X.-Z. Huang, H.-L. Zhang, L. He, X.-Y. Chang, and L.-M.
Duan, Phys. Rev. B {\bf 98}, 064306 (2018).

\bibitem{peng-2018} S. Peng, X. Xu, K. Xu, P. Huang, P. Wang, X. Kong, X. Rong, F.
Shi, C. Duan, and J. Du, Sci. Bull. {\bf 63}, 336 (2018).

\bibitem{haase-2018} J. F. Haase, P. J. Vetter, T. Unden, A. Smirne, J. Rosskopf, B.
Naydenov, A. Stacey, F. Jelezko, M. B. Plenio, and S. F. Huelga,
Phys. Rev. Lett. {\bf 121}, 060401 (2018).

\bibitem{pineda-2011} M. \v{Z}nidari\v{c}, C. Pineda, and I.
Garc\'{i}a-Mata, Phys. Rev. Lett. {\bf 107}, 080404 (2011).

\bibitem{viyuela-2012} O. Viyuela, A. Rivas, and M. A. Martin-Delgado, Phys. Rev. B {\bf 86}, 155140 (2012).

\bibitem{vasseur-2015} R. Vasseur, S. A. Parameswaran, and J. E.
Moore, Phys. Rev. B {\bf 91}, 140202(R) (2015).

\bibitem{marzolino-2017} U. Marzolino and T. Prosen, Phys. Rev. B {\bf 96}, 104402 (2017).

\bibitem{budini-2013} A. A. Budini, Phys. Rev. A {\bf 88}, 032115 (2013).

\bibitem{xue-2015} S. Xue, M. R. James, A. Shabani, V. Ugrinovskii, and I. R. Petersen, Quantum filter for a class of non-Markovian quantum systems, \emph{54th IEEE Conference on Decision and Control (Osaka, Japan)} (2015), pp. 7096-7100.

\bibitem{xue-2017} S. Xue, T. Nguyen, M. R. James, A. Shabani, V. Ugrinovskii, and I. R.
Petersen, Modelling and filtering for non-Markovian quantum
systems, arXiv:1704.00986.

\bibitem{tamascelly-2018} D. Tamascelli, A. Smirne, S. F. Huelga, and M. B. Plenio, Phys. Rev. Lett. {\bf 120}, 030402 (2018).

\bibitem{imamoglu-1994} A. Imamoglu, Phys. Rev. A {\bf 50}, 3650 (1994).

\bibitem{garraway-1997} B. M. Garraway, Phys. Rev. A {\bf 55}, 2290 (1997).

\bibitem{mazzola-2009} L. Mazzola, S. Maniscalco, J. Piilo, K.-A. Suominen, and B. M.
Garraway, Phys. Rev. A {\bf 80}, 012104 (2009).

\bibitem{gks-1976} V. Gorini, A. Kossakowski, and E. C. G. Sudarshan, J. Math. Phys. (N.Y.) {\bf 17}, 821 (1976).

\bibitem{lindblad-1976} G. Lindblad, Commun. Math. Phys. {\bf 48}, 119 (1976).

\bibitem{luchnikov-ml-2019} I. A. Luchnikov, S. V. Vintskevich, D. A. Grigoriev, and S. N.
Filippov, Machine learning of Markovian embedding for
non-Markovian quantum dynamics, arXiv:1902.07019.

\bibitem{shrapnel-2018} S. Shrapnel, F. Costa, and G. Milburn, Int. J. of Quantum Inf. {\bf 16}, 1840010 (2018).

\bibitem{Kosloff2013} R. Kosloff, Entropy {\bf 15}, 2100 (2013).

\bibitem{de-vega-2017} I. de Vega and D. Alonso, Rev. Mod. Phys. {\bf 89}, 015001 (2017).

\bibitem{arnoldus-1987} H. F. Arnoldus and T. F. George, J. Math. Phys. (N. Y.) {\bf 28}, 2731 (1987).

\bibitem{weiss} U. Weiss, {\it Quantum Dissipative Systems}, 2nd ed. (World
Scientific, Singapore, 1999).

\bibitem{eckel-2006} M. Thorwart, J. Eckel, and E. R. Mucciolo, Phys. Rev. B {\bf 72}, 235320
(2005).

\bibitem{strathearn-2018} A. Strathearn, P. Kirton, D. Kilda, J. Keeling, and B. W. Lovett, Nat. Commun. {\bf 9}, 3322 (2018).

\bibitem{Suzuki1985} M. Suzuki, J. Math. Phys. {\bf 26}, 601 (1985).

\bibitem{chiribella-2008} G. Chiribella, G. M. D'Ariano, and P. Perinotti, Phys. Rev. Lett.
{\bf 101}, 060401 (2008).

\bibitem{chiribella-2009} G. Chiribella, G. M. D'Ariano, and P. Perinotti, Phys.
Rev. A {\bf 80}, 022339 (2009).

\bibitem{hardy-2012} L. Hardy, Phil. Trans. R. Soc. A {\bf 370},
3385 (2012).

\bibitem{milz-2017} S. Milz, F. A. Pollock, and K. Modi, Open
Syst. Inf. Dyn. {\bf 24}, 1740016 (2017).

\bibitem{costa-2016} F. Costa and S. Shrapnel, New J. Phys. {\bf 18},
063032 (2016).

\bibitem{makri-1995} N. Makri and D. E. Makarov, J. Chem. Phys. {\bf 102}, 4611
(1995).

\bibitem{sim-2001} E. Sim, J. Chem. Phys. {\bf 115}, 4450 (2001).

\bibitem{strathearn-2017} A. Strathearn, B. W. Lovett, and P. Kirton, New J. Phys. {\bf
19}, 093009 (2017).

\bibitem{pollock-2019} M. R. J{\o}rgensen and F. A. Pollock, Exploiting the causal
tensor network structure of quantum processes to efficiently
simulate non-Markovian path integrals, arXiv:1902.00315.

\bibitem{pollock-2018} F. A. Pollock, C. Rodr\'{i}guez-Rosario, T. Frauenheim, M. Paternostro, and K.
Modi, Phys. Rev. A {\bf 97}, 012127 (2018).

\bibitem{modi-2018} F. A. Pollock, C. Rodr\'{i}guez-Rosario, T. Frauenheim, M. Paternostro, and K.
Modi, Phys. Rev. Lett. {\bf 120}, 040405 (2018).

\bibitem{milz-2018} S. Milz, F. A. Pollock, and K. Modi, Phys. Rev. A {\bf 98}, 012108 (2018).

\bibitem{feynman-1963} R. P. Feynman and F. L. Vernon, Ann. Phys. (N. Y.) {\bf 24}, 118 (1963).

\bibitem{Verstraete2006} F. Verstraete and J. I. Cirac, Phys. Rev. B {\bf 73}, 094423 (2006).

\bibitem{supmat}See Supplemental Material for details, which includes Refs. \cite{sason-2018,nielsen-2001,nielsen-2000}.

\bibitem{sason-2018} I. Sason, Entropy {\bf 20}, 896 (2018).

\bibitem{nielsen-2001} M. A. Nielsen and G. Vidal, Quantum Inf. Comput. {\bf 1}, 76 (2001).

\bibitem{nielsen-2000} M. A. Nielsen and I. L. Chuang, {\it Quantum Computation and
Quantum Information} (Cambridge University Press, Cambridge, England, 2000).

\bibitem{lorenzo-2017} S. Lorenzo, F. Ciccarello, and G. M. Palma, Phys. Rev. A, {\bf 96}, 032107 (2017).

\bibitem{MERA} G. Vidal, Phys. Rev. Lett. {\bf 101}, 110501 (2008).

\bibitem{Evenbly2014} G. Evenbly and G. Vidal, Phys. Rev. Lett. {\bf 112}, 240502 (2014).

\bibitem{Carleo2017} G. Carleo and M. Troyer, Science {\bf 355}, 602 (2017).

\bibitem{Amin2018} M. H. Amin, E. Andriyash, J. Rolfe, B. Kulchytskyy, and R. Melko, Phys. Rev. X {\bf 8}, 021050 (2018).

\end{thebibliography}
\end{document}